\pdfoutput=1
\documentclass[3p,times,twocolumn]{elsarticle}
\usepackage{tikz}
\usepackage{hyperref}

\usepackage{amssymb}
\usepackage{amsmath}

\biboptions{sort&compress}

\newcommand{\mpp}{m_{\pi^\pm}}
\newcommand{\mpn}{m_{\pi^0}}
\let\today\relax

\begin{document}
\begin{frontmatter}
\title{Determination of the $m_u$ and $m_d$ quark masses from $\eta\to3\pi$ decay\tnoteref{t1}}
\tnotetext[t1]{This work is supported by the Center for Particle Physics (project no.\ LC 527) of the Ministry of Education of the Czech Republic.\\ 
The presented results of the $\eta\to3\pi$ analysis stem from the work done in collaboration with K.~Kampf, M.~Knecht and J.~Novotn\'{y}.}

\author{Martin Zdr\'ahal\corref{c1}}
\cortext[c1]{\emph{Email address: }\tt zdrahal@ipnp.mff.cuni.cz}

\address{Institute of Particle and Nuclear Physics, Faculty of Mathematics and Physics, Charles University,
V Hole\v{s}ovi\v{c}k\'ach 2, Prague, Czech Republic\\[-1pt]
}

\begin{abstract}
For a precise determination of the $m_u$ and $m_d$ quark masses, it is useful to combine isospin symmetric results of lattice or sum rules QCD techniques with some isospin breaking study performed in chiral perturbation theory (ChPT).
The most promising process for the later is $\eta\to3\pi$ decay. However, this process is affected by large chiral corrections and there are observed discrepancies between the values of Dalitz plot parameters stemming from the standard ChPT computation of its amplitude and those experimentally measured. We describe here the method based on analytic dispersive representation which attempts to obtain the information on the masses by taking this discrepancy into account independently of its exact origin.
Together with the results of this analysis we present a review of some other constraints on the light quark masses and conclude with values of the masses compatible with them.
\end{abstract}


\end{frontmatter}

\section{Introduction}\label{introduction}
Quark masses are fundamental free parameters of the Standard model. Because of the QCD phenomenon of color confinement, quarks are bounded inside hadrons and their masses cannot be directly measured. Instead, for their determination one uses some physical observable depending on the masses in a theoretical approach and compares such theoretical prediction with the corresponding experimental value. Such definition of the masses thus depends on the theoretical approach used. In the following we deal with the \emph{current quark masses}, which are the masses appearing in the QCD Lagrangian. At low-energy region, where the color confinement reigns, the non-perturbative methods of QCD are required. Those relevant for the determination of the light quark masses ($m_u, m_d$, $m_s$) are sum rules (SR), lattice (LQCD) and effective field theories (EFT).

QCD sum rules are based on dispersion relations stemming from analytic properties of some observable (such as differential decay rate of $\tau$). They connect together the hadron world measured in experiments with perturbative QCD in terms of operator product expansion. For the recent review on the SR results on light quark masses we refer to \cite{PDG,Dominguez:2011sb}. Note that a computation of electromagnetic (EM) corrections within SR methods would be very difficult since the inclusion of long-range interactions changes significantly analytic properties of the amplitudes, which makes it impossible to write dispersion relations in the regular form.

Numerical simulations on lattice have achieved in recent years a considerable progress --- the current day simulations are performed with $2+1$ dynamical quarks, moreover, with $m_s$ near its physical value, while in order to reach the physical point in $m_{u,d}$ chiral extrapolation is predominantly still necessary.
However, again the inclusion of EM interactions on the lattice is very involved for their long-range character. Nevertheless, even in this aspect, there is some progress as quenched QED simulations are already performed (e.g.~\cite{Blum:2010ym}). In recent years there appeared two attempts \cite{Colangelo:2010et, Laiho:2009eu} to review and average existing lattice calculations of various quantities, including $m_q$, where more details can be found.

Both of these methods independently determined the values of $m_s$ and of isospin averaged $\hat{m}=\frac{m_u+m_d}{2}$
with a reasonable precision with more-or-less compatible results. But since the isospin breaking effects on the observables studied by both these methods generated by EM interactions are of the same order as those stemming from $m_u-m_d$ difference, for determination of the individual $m_u$,  $m_d$ they both need additional input\footnote{Even for a precise determination of $m_s$, EM interactions cannot be neglected. However, since they are here less important, at the current level of precision of $m_s$ it is enough to take an estimate of them.}.

A method that can provide such input is chiral perturbation theory (ChPT) \cite{Gasser:1984gg}, playing a prominent role among those EFT. Note that for the determination of the quark masses ChPT alone is insufficient. As in all physical results $m_q$ occurs multiplied by scalar quark condensate $B_0$ and thus rescalling both of them does not change the physics. Moreover, starting at next-to-leading order (NLO) in chiral counting, there exists a transformation of the masses \cite{Kaplan:1986ru}, which together with the corresponding change in low-energy constants $L_6, L_7, L_8$ and some of NNLO constants $C_i$ leave pseudoscalar masses, scattering amplitudes and matrix elements containing vector or axial vector currents invariant. Thus, using only experimental measurements we cannot fix this so-called Kaplan-Manohar (KM) ambiguity. Consequently, ChPT can determine only quark mass ratios and needs some external theoretical input in order to fix the physical definition of the masses. For a review of ChPT NLO determinations of light quark masses, we refer to \cite{Leutwyler:2009jg}. In addition, in \cite{Amoros:2001cp} the employment of NNLO meson mass formulae is included.

For a more precise determination of light quark masses it is therefore useful to combine isospin symmetric results of LQCD and SR with some isospin-breaking study performed in ChPT. A very suitable process for such study is $\eta\to3\pi$ decay, which is possible only in the isospin breaking world. Moreover, EM contributions to this decay are very small and thus its amplitude is to a good approximation directly proportional to $m_d - m_u$. (For a more detailed introduction to this process see \cite{eta-phenom}.) We pull out the following normalization factor out of its amplitude
\begin{equation}
\mathcal{A}(s,t,u)=\frac{\sqrt{3}}{4R}M(s,t,u),\quad \text{with}\ \  R=\frac{m_s - \hat{m}}{m_d - m_u}\,.
\end{equation}
By computation of so defined $M(s,t,u)$ in ChPT and comparing it to measured decay rate of this decay $\Gamma$, one can determine the parameter $R$. After pulling out this isospin breaking parameter we can perform isospin limit in the rest of the amplitude $M(s,t,u)$, which is a very good (but once the experimental data on $\eta\to3\pi$ decays are available with a good precision, unnecessary) approximation. In this approximation the amplitudes of the neutral decay $\eta\to3\pi^0$ and of the charged one $\eta\to\pi^+\pi^-\pi^0$ are related together. In the following, we therefore work in such first order in the isospin breaking violation (or in other words in the limit $\mpp=\mpn$) and deal mainly with the charged $\eta$ decay.

\section{\texorpdfstring{$\eta\to3\pi$}{[eta] -> 3[pi]} in ChPT}
Quite recently, an inclusion of the two-loop corrections to this decay amplitude was performed in \cite{BG}.

However, from the computed three successive orders it is obvious that chiral corrections in this process are large --- we illustrate this on listing the values of $R$ stemming from them and experimentally measured $\Gamma$ \cite{PDG}:
\begin{equation}
R_{\mathrm{LO}}=19.1,\quad R_{\mathrm{NLO}}=31.8, \quad R_{\mathrm{NNLO}}=41.3.
\end{equation}

Moreover, although the NNLO computation \cite{BG} leads to a reasonable result for the amplitude and for the value of $R$, there are still discrepancies between the experimentally measured values of Dalitz parameters describing their energy dependencies and the values predicted from \cite{BG}. The charged amplitude is usually parameterized in the form (normalized to one at the center of Dalitz plot $s=t=u=\bar{s}=\frac13(m_\eta^2+3m_\pi)$).
\begin{equation}
\frac{|M_{x}(x,y)|^2}{|M_{x}(0,0)|^2}=1+a y+b y^2+ d x^2 + f y^3+g x^2 y+\dots
\end{equation}
with variables $x=\frac{\sqrt3 (u-t)}{2 m_\eta (m_\eta-3m_\pi)}\,,\ y=\frac{3(\bar{s} -s)}{2m_\eta (m_\eta-3m_\pi)}\,.$

The parametrization of the neutral decay reads
\begin{equation}
\frac{|M_0(x,y)|^2}{|M_0(0,0)|^2}=1+2\alpha z+2\beta y(3z-4y^2)+\dots,
\end{equation}
where in addition to the variable $y$, there appears distance from the center of the Dalitz plot $z=x^2+y^2$.

\begin{table}
\centering
\begin{tabular}{|c|@{$\,$}r@{$\:\pm\:$}l@{$\;\;\;$}r@{$\:\pm\:$}l@{$\;\;$}|}\hline
&\multicolumn{2}{c}{KLOE \cite{KLOE}} & \multicolumn{2}{c|}{ChPT \cite{BG}}\\\hline
$a$
& $-1.09$  & $0.02$
& $-1.271$  &  $0.075$
\\
$b$
&{$0.124$} &  {$0.012$}
&{$0.394$}  & {$0.102$}
\\
$d$
& $0.057$ & $0.017$
& $0.055$  &  $0.057$
\\
$f$
& $0.14$  & $0.02$
& $0.025$  &  $0.160$
\\
$g$
& \multicolumn{2}{c}{$\sim0$}
& \multicolumn{2}{c|}{$0$}
\\\hline
$\alpha$
&  {$-0.030$} & {$0.005$}
&  {$0.013$}  & {$0.016$}
\\\hline
\end{tabular}
\caption{Comparison of the experimentally measured values of Dalitz plot parameters with the values stemming from NNLO ChPT}
\label{Dalitz parameters}
\end{table}

In Table~\ref{Dalitz parameters} we list the values of Dalitz parameters stemming from NNLO ChPT \cite{BG} and their experimental determination from KLOE \cite{KLOE} as the only measurement giving all the quoted parameters with reasonable precision. Note the discrepancy of the central values of parameters $b$ and $\alpha$.

A third complication in the computation as well as a possible explanation of the discussed discrepancies in Dalitz parameters is the poor knowledge of the $O(p^6)$ low-energy constants $C_i$. Two-loop amplitudes depend on subsets of 102 $C_i$s, whose determination is needed before any reliable prediction, but nowadays many of the $C_i$s are just estimated (predominantly from resonance saturation).

These complications impose the following questions: What is the origin of the Dalitz parameters discrepancy? How do this discrepancy together with the slow chiral convergence and the poor knowledge of the $C_i$s influence the determination of the isospin breaking ratio $R$?

There exist various alternative approaches taking different assumptions than standard ChPT \cite{KWW,Anisovich:1996tx, Colangelo:2009db,Schneider:2010hs,Kolesar}, which investigate further possible explanations of the discrepancy, namely higher order final state rescatterings, an influence of a slow convergence of $\pi\pi$ scattering or even of the $\eta\to3\pi$ amplitude itself. Unfortunately, the current situation does not enable to draw strong conclusions on these questions since we only have a set of 5 numbers from just one experiment for the charged decay and just one number (although independently confirmed by various experiments) for the neutral decay. Moreover, with the exception of the resummed ChPT study \cite{Kolesar} (which nevertheless gives only qualitative answers) these alternative approaches do not fix normalization of the amplitudes (there do not appear $m_q$ explicitly). Thus, in order to provide any information on the quark masses, they unavoidably need to be matched to ChPT, which brings into such determination additional assumptions about the region where both approaches are compatible and give physical results. (Moreover, at the time being, they deal with the problem of the poor knowledge of $C_i$s by using NLO ChPT results only.)

In \cite{eta-phenom} we have also touched the question of possible explanations of the Dalitz plot discrepancy but concentrated more on the second question and presented a method that tries to extract as much information from the experiment and the two-loop ChPT result taking into account these three complications no matter what is the exact explanation of the Dalitz parameters discrepancy.

For this purpose we have used our analytic dispersive parametrization, which is based on basic assumptions of QFT together with some hierarchy of various contributions to the amplitude (inspired by a very basic chiral counting and/or numerical studies). It takes into account two final-state rescatterings and can include full isospin violation $m_{\pi^\pm}\neq m_{\pi^0}$. In the first order of isospin breaking the charged amplitude is parametrized in the form
\begin{equation}\label{nase}
M_{x}(s,t,u)=P(s,t,u)+U(s,t,u),
\end{equation}
where $P(s,t,u)$ is the polynomial part of the amplitude
\begin{multline}\label{polynom}
P=A_x M_\eta^2+B_x(s-\bar{s})+C_x(s-\bar{s})^2+E_x(s-\bar{s})^3\\
+D_x[(t-\bar{s})^2+(u-\bar{s})^2]+F_x[(t-\bar{s})^3+(u-\bar{s})^3]
\end{multline}
and unitarity part $U$ contains the parameters $A_x,B_x,$ $C_x,D_x$ together with subthreshold parameters $\alpha_\pi,\beta_\pi,$ $\lambda_1,\lambda_2$ describing $\pi\pi$ scattering (fixed from \cite{Knecht:1995tr}).

Additional advantage of parametrization (\ref{nase}) of the amplitude is that by setting its parameters to their particular values, it can reproduce the NNLO ChPT amplitude exactly (on the region below the $\pi\eta$ thresholds).

We have presented two analyses using different assumptions on the physical amplitude trying to use the information about it stemming from NNLO ChPT together with the one from KLOE. As we have already noted, since the original data sets of KLOE are inaccessible, we had to construct a distribution from the values of five Dalitz parameters published by them (including their errors) and all our current analyses depend on the assumption that such distribution describes well the genuine physical amplitude, thereby relying on KLOE error determination.

\section{First analysis: Correcting ChPT \texorpdfstring{$\eta\to3\pi$}{[eta] -> 3[pi]} result}
Our first analysis is inspired by the following observation. At the two-loop level, one can find some relations between Dalitz parameters that are independent on the values of the $C_i$s. They read
\begin{enumerate}
\setlength{\labelsep}{0mm}
\item[rel$_1$:] $\ \big(4(b+d)-a^2-16\alpha\big)\big\vert_C=0$
\item[rel$_2$:] $\ \big(a^3-4ab+4ad+8f-8g\big)\big\vert_C=0\quad$\hfill (CIR)
\item[rel$_3$:] $\ \beta\vert_C=0$.
\end{enumerate}
If we compare the values of these combinations coming from KLOE with those from NNLO ChPT computation, we find a good correspondence between their central values, as is obvious from Table~\ref{tabulka:CIR}. (For comparison we have also included there the values corresponding to the alternative approach of \cite{Schneider:2010hs} (NREFT), where a possible explanation for the $\alpha$ discrepancy is proposed.) 

It motivates us to assume in this analysis that all the discrepancy can be included into a small real polynomial contribution (i.e.~into a possible change of the values of the $C_i$s).

\begin{table}
\centering
\begin{tabular}{|c|r@{$\:\pm\:$}lr@{$\:\pm\:$}lr@{$\:\pm\:$}l|}\hline
&\multicolumn{2}{c}{KLOE \cite{KLOE}} & \multicolumn{2}{c}{ChPT \cite{BG}}& \multicolumn{2}{c|}{NREFT \cite{Schneider:2010hs}}\\\hline
rel$_1$
& $0.02$  &  $0.12$
& $-0.03$  &  $0.72$
& $0.35$  &  $0.13$
\\
\hline
rel$_2$
& $0.12$  &  $0.21$
& $-0.13$  &  $1.4$
& $0.44$  &  $0.20$
\\
\hline
$10^3\beta$
& \multicolumn{2}{c}{?}
& $-2$  &  $25$
& $-4.2$  &  $0.7$
\\
\hline
\end{tabular}
\caption{Values of $C_i$-independent combinations (CIR) of Dalitz parameters corresponding to various determinations.}\label{tabulka:CIR}
\end{table}

Before we proceed to such analysis let us make a few comments. The error bars quoted in Table~\ref{tabulka:CIR} are overestimated since they were obtained just as combinations of the errors of the individual parameters --- this is the case especially for the ChPT values, where the parameters were affected by the large $C_i$ uncertainties, whose effect on these combinations is reduced. Hence, it would be desirable to remeasure these combinations (and reevaluate them also in the ChPT analysis) in order to test this class of explanations. A case of a special importance is the neutral parameter $\beta$, which is thus determined in ChPT independently on the $C_i$s. 

Once this correspondence is confirmed, from the values of the individual Dalitz parameters [or more suitably of those of parametrization (\ref{nase})] corresponding to the experimental amplitude one could construct sum rules for the $C_i$s and examine their compatibility with the values coming from other observables.

Now let us return to our analysis. We assume that by an addition of a small real polynomial $\Delta P$ to the NNLO ChPT amplitude one would reproduce the physical data. By fitting such $\Delta P$ from KLOE we obtain such ``corrected amplitude''. We have found \cite{eta-phenom} that the corrections are indeed small (on the physical region) and correspond to the change of parameters in polynomial part (\ref{polynom}), which is shown in Table~\ref{tabulka:prvni}. Even without changing its unitarity part, $M_{\mathrm{corr.}}$ reproduces very well the KLOE-like distribution and the corresponding value of $R$ is shifted to
\begin{equation}
R=37.7\pm 2.9 \quad \text{[ChPT+Disp.+KLOE]}.
\end{equation}
The quoted error is estimated conservatively taking into account the slow convergence of chiral expansion in the first three orders --- assuming that the error of the corrected amplitude is equal to one half of the difference between this amplitude and the NLO one (similarly as it was done with the NNLO amplitude in \cite{BG}).
\begin{table}
\centering
{\small\setlength{\tabcolsep}{0.58mm}
\begin{tabular}{|c|c|c|c|c|c|c|}\hline
 & $A$ & $B$ & $C$ & $D$ & $E$ & $F$ \\
 \hline
$P^{(4)}$ & $0.46(1)$ & $1.95(10)$ & $-0.6(2)$ & $1.04(2)$ &  &  \\
$P^{(6)}$ & $0.58(1)$ & $2.4(2)$ & $0.3(34)$ & $1.6(24)$ & $5(150)$ & $-4(84)$ \\
$P^{(6)}_{\mathrm{corr.}}$ & $0.575(6)$ & $1.99(4)$ & $-6.8(3)$ & $0.94(3)$ & $-31(3)$ & $20(1)$\\
\hline
\end{tabular}}
\caption{The values of parameters of (\ref{nase}) corresponding to the ``corrected amplitude'' of the first analysis.}\label{tabulka:prvni}
\end{table}

\section{Second analysis: Direct fit to \texorpdfstring{$\eta\to3\pi$}{[eta] -> 3[pi]} data}
A different analysis not imposing anything about the origin of the Dalitz parameters discrepancy takes advantage of the fact that the construction of parametrization (\ref{nase}) employs just general properties of the amplitude. We can thus assume that the physical amplitude (for now the KLOE-like distribution) should be reproduced by this parametrization, which gives us a clear prescription for its fitting. In order to fix the normalization we suppose that ChPT determination of the amplitude is reliable at least in the region\footnote{Note that since we have analytic results, for fixing of the normalization one matching point would be sufficient.} specified below.

The usually employed fact when searching for such matching region is the accidental coincidence at the chiral NLO level of the following points (on the cut $s=u$):
\begin{enumerate}
 \item[a)] the point, where the amplitude is of order $O(m_\pi^2)$ (Adler zero),
 \item[b)] the point, where $\mathrm{Re}\,M=0$,
 \item[c)] the point, where the corrections of the computed order to the slope are small.
\end{enumerate}
However, this coincidence proves not to be the case at NNLO. Therefore, having the NNLO results at hand there is no reason why the amplitude at e.g.~the point fulfilling condition b) should have a faster chiral convergence than at any other point (in this particular case at NNLO it even seems to be the opposite).

\begin{figure}
\centering
\includegraphics[width=0.9\columnwidth]{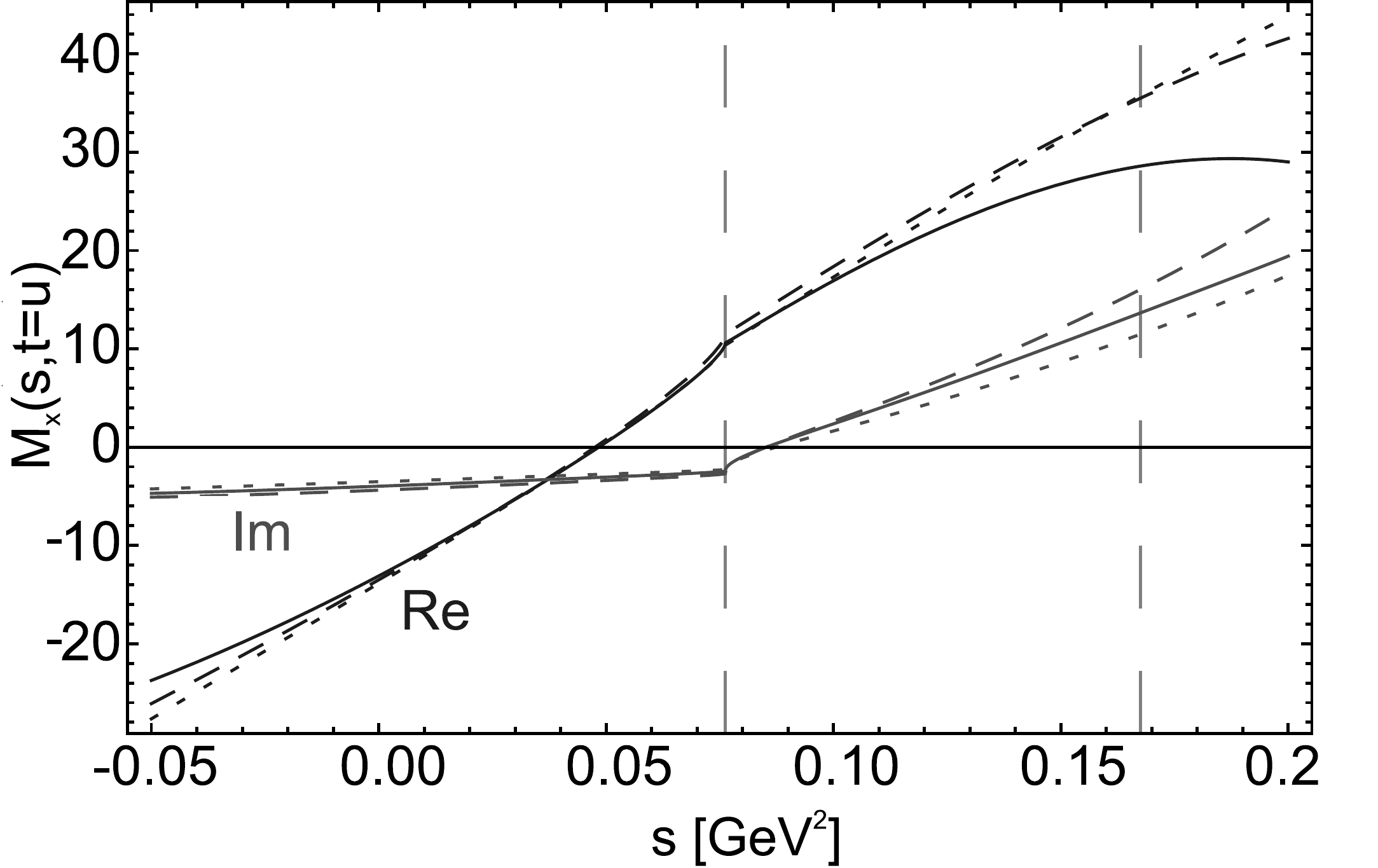}
\caption{Comparison of the amplitude obtained by our second analysis (solid lines) with the order-by-order parametrization of the NNLO ChPT amplitude (dotted) and its ``resummed'' parametrization (dashed) on the cut $t=u$. (More details in the main text.) The vertical lines indicate the physical region.}\label{Fity}
\end{figure}

Instead, taking parametrization (\ref{nase}) and the first three chiral orders of the amplitude that are at disposal, we have found a more suitable prescription\footnote{However, it could happen that in future e.g.~some NNNLO computation reveals that even this prescription is not the best one.} \cite{eta-phenom}: We should fit on the $t=u$ cut and match only imaginary parts of the amplitudes under the physical thresholds (i.e.~the dependence on the $C_i$s is reduced). Moreover, one can further reduce the error of this procedure by fixing the normalization so that the resulting parametrization in this region interpolates between the order-by-order and the ``resummed" fits (\ref{nase}) of the chiral amplitude (they differ in the unitarity part, in the choice, whether one respects chiral orders of the parameters appearing there or whether one reorders them taking into account also some higher unitarity contributions).

In Figure~\ref{Fity} we plot the result of this matching in comparison to the order-by-order and the ``resummed'' parametrization of the original NNLO ChPT amplitude.

By comparing measured $\Gamma$ \cite{PDG} with integration of the resulting amplitude over phase space, we obtain
\begin{equation}
R=37.8\pm 3.3 \quad \text{[Disp.+KLOE]}.
\end{equation}
The possible sources of errors entering this analysis include: the uncertainties of the experimental data together with the (in)accuracy of their description by our parametrization; the errors induced by analytic continuation of the parametrization to the region where we match to ChPT; and finally the error of chiral expansion of the amplitude (its $O(p^6)$ order) which is used for the matching in this region. Taking into account properties of parametrization (\ref{nase}) and the above described construction of the matching procedure, there is no surprise that the first type of errors prevails over other two even if we estimate them conservatively.

Further details on both analyses can be found in \cite{eta-phenom}.

\section{Results}
Since the dominant errors in both the performed analyses are of different origin (and the values are compatible), we can combine them into
\begin{equation}\label{vysledek}
R=37.7\pm2.2.
\end{equation}
However, as was stated above, both analyses rely on the assumption that the genuine physical amplitude is described by the distribution constructed from the values of Dalitz parameters given by KLOE (and on the fact that one can use the NNLO ChPT result at least as the input for the matching). Without further measurements, the error bar connected with this assumption is difficult to quantify. We estimate it from the second analysis and use in the following also more conservative value
\begin{equation}\label{opatrny}
R=37.7\pm3.3.
\end{equation}

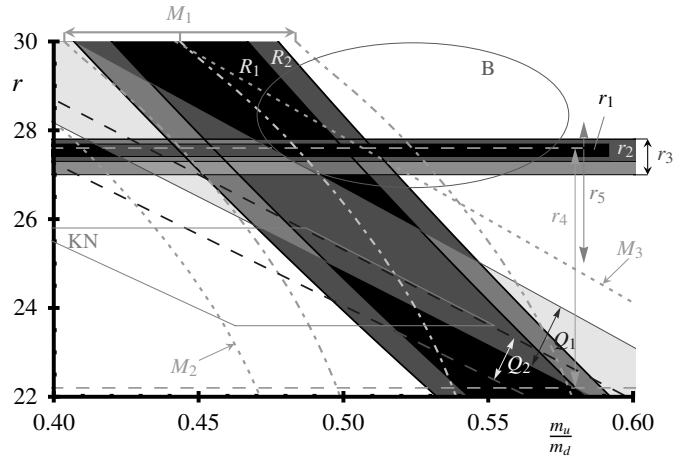
\begin{figure}
\mbox{}\\[8pt]
\includegraphics[width=\columnwidth]{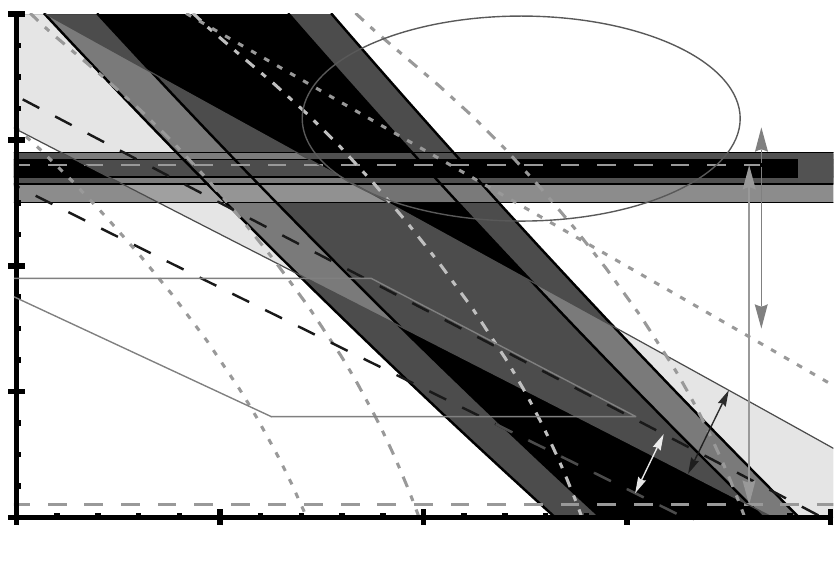}\\[0pt]
\begin{picture}(0,0)(-4.5,-24.25)
\begin{tikzpicture}[overlay,xscale=1.905,yscale=1.176,>=stealth]%
\draw (0,-0.1) node[below] {\small $0.40$};%
\draw (1,-0.1) node[below] {\small $0.45$};%
\draw (2,-0.1) node[below] {\small $0.50$};%
\draw (3,-0.1) node[below] {\small $0.55$};%
\draw (4,-0.1) node[below] {\small $0.60$};%
\draw (-0.05,0) node[left] {\small $22$};%
\draw (-0.05,1) node[left] {\small $24$};%
\draw (-0.05,2) node[left] {\small $26$};%
\draw (-0.05,3) node[left] {\small $28$};%
\draw (-0.05,4) node[left] {\small $30$};%
\draw (3.5,-0.2) node[below] {$\frac{m_u}{m_d}$};%
\draw (-0.15,3.5) node[left] {$r$};%
\draw[gray] (3.73,2.775)--(3.8,3.15);
\draw (3.82,3.15) node[above,black] {\footnotesize $r_1$};
\draw (3.95,2.765) node[white!90!black] {\footnotesize $r_2$};
\draw[black,very thin] (4,2.897)--(4.1,2.897) (4,2.493)--(4.1,2.493);
\draw[<->] (4.1,2.493)--(4.1,2.897);
\draw (4.1,2.695) node[right,white!30!black] {\footnotesize $r_3$};
\draw (3.62,2) node[left,white!60!black] {\footnotesize $r_4$};
\draw (3.63,2.25) node[right,white!50!black] {\footnotesize $r_5$};
\draw (3.06,0.35) node[right,white] {\footnotesize $Q_2$};
\draw (3.38,0.62) node[right,white!10!black] {\footnotesize $Q_1$};
\draw (1.5,3.7) node[left,white!90!black] {\footnotesize $R_1$};
\draw (1.72,3.8) node[left,white!90!black] {\footnotesize $R_2$};
\draw[white!60!black,thick] (0.071,4)--(0.071,4.1) (0.87,4)--(0.87,4.1) (1.668,4)--(1.668,4.1);
\draw[<->,white!60!black,thick] (0.071,4.1)--(1.668,4.1);
\draw (0.87,4.1) node[above,white!60!black] {\footnotesize $M_1$};
\draw[<-,white!60!black] (1.245,0.585)--(1,0.4);
\draw[white!60!black] (0.9,0.32) node {\footnotesize $M_2$};
\draw[<-,white!60!black] (3.79,1.265)--(3.97,1.55);
\draw[white!60!black] (3.98,1.65) node {\footnotesize $M_3$};
\draw[white!35!black] (3,3.7) node {\footnotesize B};
\draw[white!50!black] (0.2,1.75) node {\footnotesize KN};
\end{tikzpicture}
\end{picture}\\[-15pt]
\caption{Various constraints on quark mass ratios (description given in the main text)}\label{fig:Ratios}
\end{figure}

In Figure~\ref{fig:Ratios} we review various constraints on the light quark masses coming from various recent analyses. In order to keep the plot uncluttered, we include only those which are independent and when different updates of some analysis exist, we list only the most recent one. For older results we refer to \cite{Leutwyler:2009jg}. The axes of the presented plot correspond to (isospin symmetric) ratio $r=\frac{m_s}{\hat{m}}$ and to quantitative measure of isospin violation $\frac{m_u}{m_d}$, which appear naturally in various analyses.%
\footnote{The different plot used in previous summaries (as e.g.\ \cite{Leutwyler:2009jg}) is connected with another isospin breaking parameter $Q^2=\frac{m_s^2-\hat{m}^2}{m_d^2-m_u^2}$ which have at NLO in ChPT the advantages that it can be expressed in terms of QCD pseudoscalar masses only and is reasonably stable with respect to KM transformation. However, at the current level of precision one includes also chiral two-loop effects, with which both of these advantages of $Q$ are lost since the relation between $Q$ and the meson masses gains noticeable $r$-dependent corrections at NNLO (cf.\ \cite{Amoros:2001cp}).

Also in our analyses, instead of $M$ we could have used $M_Q$ defined by pulling out $Q^2$. However, since we need to match to NNLO ChPT \cite{BG}, the normalization with $R$ is more natural. Note that KM ambiguity is assumed to be fixed in \cite{BG} by the values of LECs used there (e.g.\ the value of $L_6^r=0$ stemming from large $N_c$ considerations).}

Our result (\ref{vysledek}) is denoted by $R_1$ whereas the more conservative one by $R_2$. The results of the alternative analysis of $\eta\to3\pi$ from \cite{Anisovich:1996tx,Colangelo:2009db} are depicted as $Q_1$ (with the error bar estimated by \cite{Leutwyler:2009jg}) and the more recent result\footnote{Talk by E.~Passemar at EPS HEP 2011, Grenoble, July 2011.} with $Q_2$. This numeric approach uses dispersive relations of Omn\`{e}s type, thereby attempting to include two-pion rescatterings to all orders. The necessary matching to ChPT is performed at the NLO Adler zero (see above) to the NLO amplitude. For $r>26$ it leads (similarly as ChPT NLO computation) to $m_u$ lighter than our result.

Interestingly, our result is fully compatible with the result of \cite{Amoros:2001cp} using ChPT NNLO expressions for meson masses that take into account only the EM contributions due to Dashen \cite{Dashen:1969eg} (central value and error bars denoted by $M_1$). The result of \cite{Amoros:2001cp} that includes the violation of Dashen limit computed in \cite{Bijnens:1996kk} $x_D=1.84$ (depicted without error bars as $M_2$) moves $m_u$ to noticeable smaller values. If one used only NLO expressions for the masses, for the value of $r\lesssim28$ our result would lie in between the determination respecting Dashen ($M_3$) and the one with $x_D=1.84$ (slightly above the upper bound of $Q_2$). For $r<27$ this NLO determination violating Dashen limit would be more compatible with our result than the one respecting it.

We proceed now to the recent determinations of $r$. Progress report on sum rules \cite{Dominguez:2011sb} quotes the value denoted by $r_4$. The method of \cite{Kaiser:1998ds} inspired by large $N_c$ considerations enables to determine $r$ from the ratio of $\frac{\Gamma(\eta\to\gamma\gamma)}{\Gamma(\eta'\to\gamma\gamma)}$. The original number is indicated by arrow $r_5$. In \cite{Kastner:2008ch} such analysis was repeated with more recent numbers, giving the only recent determination of $r$ incompatible with the lattice averages. There was also used the result of \cite{Ananthanarayan:2004qk} for Dashen violation $x_D=1.5$ and from $m_{K^+} - m_{K^0}$ obtained a different value of $Q$. These two results of \cite{Kastner:2008ch} together are depicted with label KN.

The most advanced determinations of $r$ are performed using lattice. Nowadays there exist two averaging attempts on lattice results. The average of Laiho et al.\ \cite{Laiho:2009eu} gives $r_1=27.55(14)$ and isospin quark masses\footnote{Unlike the mass ratios used above, $m_q$ depend on renormalization scheme. All quoted values are in $\overline{MS}$ scheme at the scale $\mu=2\,\mathrm{GeV}$.}
\begin{equation}
m_s^{\mu}=93.6(1.1)\,\mathrm{GeV}, \quad \hat{m}^{\mu}=3.419(47)\,\mathrm{GeV}.
\end{equation}
FLAG group \cite{Colangelo:2010et} have similar averages but in order to be more conservative, they quote estimates $r_3=27.4(4)$,
\begin{equation}\label{r_3}
m_s^{\mu}=94(3)\,\mathrm{GeV}, \quad \hat{m}^{\mu}=3.43(11)\,\mathrm{GeV}.
\end{equation}

We use FLAG numbers as our conservative choice. However, since their estimate are in some aspects very strict (e.g.\ not including BMW \cite{Durr:2010aw} yet), in order to show the precision of the current determinations we also use the average of \cite{Laiho:2009eu} with error bars extended so that they include all the central values of the individual recent lattice determinations, $r_2=27.55(25)$,
\begin{equation}\label{r_2}
m_s^{\mu}=93.6(2.2)\,\mathrm{GeV}, \quad \hat{m}^{\mu}=3.42(9)\,\mathrm{GeV}.
\end{equation}

In Figure~\ref{fig:Ratios} we have also included recent isospin breaking study on lattice \cite{Blum:2010ym} denoted by B (note that it is included in average \cite{Laiho:2009eu} but not in \cite{Colangelo:2010et}). For completeness, let us note that baryonic determinations as \cite{Guo:2010ca} give smaller quark ratio $\frac{m_u}{m_d}\lesssim0.4$.

Taking our result (\ref{opatrny}) together with the lattice estimate (\ref{r_3}), we obtain
\begin{gather*}
\frac{m_u}{m_d}=0.48(3)\ \left[0.48(2)\right],\quad Q= 23(1)\ \left[23.2(7)\right],\\
m_u^{\overline{MS},\,\mu=2\,\mathrm{GeV}}=2.23(13)\,\mathrm{MeV}\ \left[2.21(9)\,\mathrm{MeV}\right],\\
 m_d^{\overline{MS},\,\mu=2\,\mathrm{GeV}}=4.63(16)\,\mathrm{MeV}\ \left[4.62(12)\,\mathrm{MeV}\right].
\end{gather*}
The more precise values given in the square brackets correspond to the situation if our assumptions on the $\eta\to3\pi$ data and the estimates on the isospin quark masses (\ref{r_2}) were fulfilled.


\begin{thebibliography}{10}
\expandafter\ifx\csname url\endcsname\relax
  \def\url#1{\texttt{#1}}\fi
\expandafter\ifx\csname urlprefix\endcsname\relax\def\urlprefix{URL }\fi
\expandafter\ifx\csname href\endcsname\relax
  \def\href#1#2{#2} \def\path#1{#1}\fi

\bibitem{PDG}
K.~Nakamura, et~al., \href{http://pdg.lbl.gov/2010/tables/contents_tables.html}{Review of particle physics}, J.\ Phys. G37 (2010) 075021.

\bibitem{Dominguez:2011sb}
C.~Dominguez, {Quark masses in QCD: a progress report}, Mod.\ Phys.\ Lett. A26
  (2011) 691--710.
\newblock \href {http://arxiv.org/abs/1103.5864} {\path{arXiv:1103.5864}}.

\bibitem{Blum:2010ym}
T.~Blum, et~al., {Electromagnetic
  mass splittings of the low lying hadrons and quark masses from 2+1 flavor
  lattice QCD+QED}, Phys.\ Rev. D82 (2010) 094508.
\newblock \href {http://arxiv.org/abs/1006.1311} {\path{arXiv:1006.1311}}.

\bibitem{Colangelo:2010et}
G.~Colangelo, et~al., {Review
  of lattice results concerning low energy particle physics}, Eur.\ Phys.\ J.
  C71 (2011) 1695. \newblock \href {http://arxiv.org/abs/1011.4408} {\path{arXiv:1011.4408}}. Updates at \href{http://itpwiki.unibe.ch/flag}{\path{itpwiki.unibe.ch/flag}}.

\bibitem{Laiho:2009eu}
J.~Laiho, E.~Lunghi, R.~S. Van~de Water, {Lattice QCD inputs to the CKM
  unitarity triangle analysis}, Phys.Rev. D81 (2010) 034503.
\newblock \href {http://arxiv.org/abs/0910.2928} {\path{arXiv:0910.2928}}.
J.~Laiho, {Light quark physics from lattice QCD}, \href
  {http://arxiv.org/abs/1106.0457} {\path{arXiv:1106.0457}}. Updates at \href{http://www.latticeaverages.org}{\path{www.latticeaverages.org}}.

\bibitem{Gasser:1984gg}
J.~Gasser, H.~Leutwyler, {Chiral Perturbation Theory: Expansions in the Mass of
  the Strange Quark}, Nucl.\ Phys. B250 (1985) 465.

\bibitem{Kaplan:1986ru}
D.~B. Kaplan, A.~V. Manohar, {Current Mass Ratios of the Light Quarks}, Phys.\
  Rev.\ Lett. 56 (1986) 2004.

\bibitem{Leutwyler:2009jg}
H.~Leutwyler, {Light quark masses}, PoS CD09 (2009) 005.
\newblock \href {http://arxiv.org/abs/0911.1416} {\path{arXiv:0911.1416}}.

\bibitem{Amoros:2001cp}
G.~Amoros, J.~Bijnens, P.~Talavera, {QCD isospin breaking in meson masses,
  decay constants and quark mass ratios}, Nucl.\ Phys. B602 (2001) 87--108.
\newblock \href {http://arxiv.org/abs/hep-ph/0101127}
  {\path{arXiv:hep-ph/0101127}}.

\bibitem{eta-phenom}
K.~Kampf, M.~Knecht, J.~Novotn\'{y}, M.~Zdr\'{a}hal, {Analytical dispersive
  construction of $\eta\to3\pi$ amplitude: first order in isospin
  breaking}, \href {http://arxiv.org/abs/1103.0982} {\path{arXiv:1103.0982}}.

\bibitem{BG}
J.~Bijnens, K.~Ghorbani, {$\eta \to 3 \pi$ at Two Loops In Chiral Perturbation
  Theory}, JHEP 11 (2007) 030.
\newblock \href {http://arxiv.org/abs/0709.0230} {\path{arXiv:0709.0230}}.

\bibitem{KLOE}
F.~Ambrosino, et~al., {Determination of $\eta\to\pi^+\pi^-\pi^0$ Dalitz plot
  slopes and asymmetries with the KLOE detector}, JHEP 05 (2008) 006.
\newblock \href {http://arxiv.org/abs/0801.2642} {\path{arXiv:0801.2642}}. F.~Ambrosino, et~al., {Measurement of the $\eta\to 3\pi^{0}$ slope parameter
  $\alpha$ with the KLOE detector}, Phys.\ Lett. B694 (2010) 16--21.
\newblock \href {http://arxiv.org/abs/1004.1319} {\path{arXiv:1004.1319}}.

\bibitem{KWW}
J.~Kambor, C.~Wiesendanger, D.~Wyler, {Final State Interactions and
  Khuri-Treiman Equations in $\eta\to 3\pi$ decays}, Nucl.\ Phys. B465 (1996)
  215--266.
\newblock \href {http://arxiv.org/abs/hep-ph/9509374}
  {\path{arXiv:hep-ph/9509374}}.

\bibitem{Anisovich:1996tx}
A.~V. Anisovich, H.~Leutwyler, {Dispersive analysis of the decay
  $\eta\to3\pi$}, Phys.\ Lett. B375 (1996) 335--342.
\newblock \href {http://arxiv.org/abs/hep-ph/9601237}
  {\path{arXiv:hep-ph/9601237}}.

\bibitem{Colangelo:2009db}
G.~Colangelo, S.~Lanz, E.~Passemar, {A New Dispersive Analysis of
  $\eta\to3\pi$}, PoS CD09 (2009) 047.
\newblock \href {http://arxiv.org/abs/0910.0765} {\path{arXiv:0910.0765}}.

\bibitem{Schneider:2010hs}
S.~P. Schneider, B.~Kubis, C.~Ditsche, {Rescattering effects in $\eta\to3\pi$
  decays}, JHEP 1102 (2011) 028.
\newblock \href {http://arxiv.org/abs/1010.3946} {\path{arXiv:1010.3946}}.

\bibitem{Kolesar}
  M.~Koles\'{a}r, {Analysis of discrepancies in Dalitz plot parameters in $\eta\to3\pi$ decay},
\href {http://arxiv.org/abs/1109.0851} {\path{arXiv:1109.0851}}.

\bibitem{Knecht:1995tr}
M.~Knecht, B.~Moussallam, J.~Stern, N.~H. Fuchs, {The Low-energy $\pi\pi$
  amplitude to one and two loops}, Nucl.\ Phys. B457 (1995) 513--576.
\newblock \href {http://arxiv.org/abs/hep-ph/9507319}
  {\path{arXiv:hep-ph/9507319}}.

\bibitem{Dashen:1969eg}
R.~F. Dashen, {Chiral SU(3) x SU(3) as a symmetry of the strong interactions},
  Phys.\ Rev. 183 (1969) 1245--1260.

\bibitem{Bijnens:1996kk}
J.~Bijnens, J.~Prades, {Electromagnetic corrections for pions and kaons: Masses
  and polarizabilities}, Nucl.\ Phys. B490 (1997) 239--271.
\newblock \href {http://arxiv.org/abs/hep-ph/9610360}
  {\path{arXiv:hep-ph/9610360}}.

\bibitem{Kaiser:1998ds}
R.~Kaiser, H.~Leutwyler, {Pseudoscalar decay constants at large N(c)}, \href
  {http://arxiv.org/abs/hep-ph/9806336} {\path{arXiv:hep-ph/9806336}}.

\bibitem{Kastner:2008ch}
A.~Kastner, H.~Neufeld, {The $K_{\ell3}$ scalar form factors in the standard
  model}, Eur.\ Phys.\ J. C57 (2008) 541--556.
\newblock \href {http://arxiv.org/abs/0805.2222} {\path{arXiv:0805.2222}}.

\bibitem{Ananthanarayan:2004qk}
B.~Ananthanarayan, B.~Moussallam, {Four-point correlator constraints on
  electromagnetic chiral parameters and resonance effective Lagrangians}, JHEP
  0406 (2004) 047.
\newblock \href {http://arxiv.org/abs/hep-ph/0405206}
  {\path{arXiv:hep-ph/0405206}}.

\bibitem{Durr:2010aw}
S.~Durr, et~al., {Lattice QCD at the
  physical point: Simulation and analysis details}, \href
  {http://arxiv.org/abs/1011.2711} {\path{arXiv:1011.2711}}.

\bibitem{Guo:2010ca}
F.-K. Guo, C.~Hanhart, U.-G. Mei\ss{}ner, {Extracting the light quark mass
  ratio $m_u/m_d$ from bottomonia transitions}, Phys.\ Rev.\ Lett. 105 (2010)
  162001.
\newblock \href {http://arxiv.org/abs/1007.4682} {\path{arXiv:1007.4682}}.

\end{thebibliography}
\end{document}